\documentclass[sigconf]{acmart}
\usepackage{algorithm}
\usepackage{algorithmic}
\usepackage{graphicx}
\usepackage{multirow}
\usepackage{makecell}

\AtBeginDocument{
  }

\copyrightyear{2025} 
\acmYear{2025} 
\setcopyright{acmlicensed}\acmConference[ICAIF '25]{6th ACM International
 Conference on AI in Finance}{November 15--18, 2025}{Singapore, Singapore}
\acmBooktitle{6th ACM International Conference on AI in Finance (ICAIF '25),
November 15--18, 2025, Singapore, Singapore}
\acmDOI{10.1145/3768292.3770415}
\acmISBN{979-8-4007-2220-2/2025/11}
\begin{document}
\title{Scaling Conditional Autoencoders for Portfolio Optimization via Uncertainty-Aware Factor Selection}

\author{Ryan Engel}
\email{Ryan.m.engel@stonybrook.edu}
\affiliation{
    \institution{Stony Brook University}
    \department{Department of Computer Science}
    \country{USA}
}

\author{Yu Chen}
\email{Yu.Chen.7@stonybrook.edu}
\affiliation{
    \institution{Stony Brook University}
    \department{Department of Applied Mathematics and Statistics}
    \country{USA}
}

\author{Pawel Polak}
\email{Pawel.Polak@stonybrook.edu}
\affiliation{
    \institution{Stony Brook University}
    \department{Department of Applied Mathematics and Statistics}
    \country{USA}
}

\author{Ioana Boier}
\email{iboier@nvidia.com}
\affiliation{
    \institution{NVIDIA Corporation}
    \country{USA}
}

\renewcommand{\shortauthors}{Engel, Chen, Polak, and Boier}

\begin{abstract}
   Conditional Autoencoders (CAEs) offer a flexible, interpretable approach for estimating latent asset-pricing factors from firm characteristics. However, existing studies usually limit the latent factor dimension to around $K=5$ due to concerns that larger $K$ can degrade performance. To overcome this challenge, we propose a scalable framework that couples a high-dimensional CAE with an uncertainty-aware factor selection procedure. We employ three models for quantile prediction: zero-shot Chronos, a pretrained time-series foundation model (ZS-Chronos), gradient-boosted quantile regression trees using XGBoost and RAPIDS (Q-Boost), and an I.I.D bootstrap-based sample mean model (IID-BS). For each model, we rank factors by forecast uncertainty and retain the top-$\kappa$ most predictable factors for portfolio construction, where $\kappa$ denotes the selected subset of factors. This pruning strategy delivers substantial gains in risk-adjusted performance across all forecasting models. Furthermore, due to each model's uncorrelated predictions, a performance-weighted ensemble consistently outperforms individual models with higher Sharpe, Sortino, and Omega ratios.
\end{abstract}



\begin{teaserfigure}
  \includegraphics[width=\textwidth]{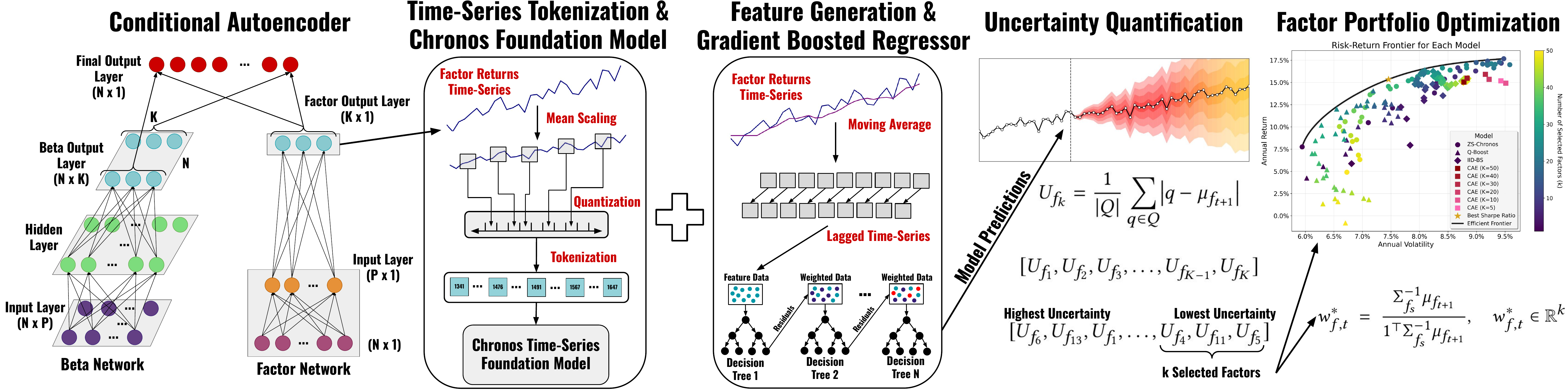}
  \caption{Architecture for Uncertainty-Aware Factor Selection. First, a CAE model extracts high-dimensional latent factor portfolios from firm characteristics. Next, time-series forecasting models generate point and quantile predictions for each factor, which are used for uncertainty quantification. Factors are then ranked by uncertainty, and the most predictable subset is selected for tangency portfolio optimization in factor space before projection to tradable asset weights.}
  \label{fig:teaser}
\end{teaserfigure}

\maketitle

\section{Introduction}

Understanding the cross-section of asset returns remains a central problem in empirical finance. Classical linear factor models—such as the Arbitrage Pricing Theory (APT) or Fama–French specifications—assume that asset returns are driven by a small number of latent or observable risk factors with fixed loadings across firms. While such models are theoretically elegant and provide economic interpretability, they rely on strong assumptions about linearity and stationarity, and often ignore the rich heterogeneity present in firm characteristics and evolving market regimes. Recent studies in \cite{neuhierl2022structural}, \cite{giglio2021asset} have highlighted limitations of these approaches, especially their inability to capture dynamic, nonlinear dependencies or time-varying exposures.

To overcome these limitations, we build upon the CAE model proposed in~\cite{gu2020autoencoder}, which generalizes linear factor models by allowing the factor loadings to depend nonlinearly on firm-specific lagged characteristics. The CAE jointly learns both the latent factors and characteristic-based exposures through a neural network architecture, enabling a more expressive representation of the return-generating process. This structure enhances the signal-to-noise ratio of latent factor returns and improves their out-of-sample predictive stability. Recent advances in deep latent factor models confirm the importance of modeling interactions between firm-level variables and latent risks, especially in high-dimensional financial environments \cite{wei2025deeplatent, neuhierl2022structural}.

A central modeling choice in CAEs concerns the dimensionality \( K \) of the latent factor space. Conventional implementations typically restrict \( K \) to small values (e.g., \( K \approx 5 \)) to maintain interpretability and limit overfitting. In contrast, we operate in a high-dimensional regime (up to \( K = 50 \)) and introduce a post-hoc filtering procedure that selects latent portfolios according to their out-of-sample predictability. For each latent factor, multiple forecasting models produce predictive distributions from which we quantify forecast uncertainty. Factors are subsequently ranked by their predictive stability, and only those with the most reliable forecasts are retained. This uncertainty-aware selection leverages the expressive capacity of high-dimensional latent representations while controlling estimation risk. By ranking factors based on predictive confidence, the procedure explicitly minimizes expected out-of-sample utility loss and thereby aligns factor selection with the downstream goal of portfolio optimization—contrasting with traditional approaches that determine \( K \) through purely statistical reconstruction criteria.

Our contributions are both methodological and empirical. First, we introduce a novel uncertainty-aware framework for factor portfolio selection, integrating high-dimensional latent portfolios extracted via CAEs with predictive signals from multiple advanced forecasting models. Second, we show that high-dimensional CAEs, when coupled with forecast-driven factor selection, significantly outperform conventional low-dimensional factor models, achieving notably higher Sharpe, Sortino, and Omega ratios while maintaining maximum drawdowns below 10\%. Third, we demonstrate that forecasts generated by zero-shot pretrained models such as Chronos \cite{ansari2024chronos} and quantile gradient-boosted regression trees provide complementary predictive signals that substantially augment the baseline IID-based predictions typically used in CAE-based frameworks. Consequently, ensembles of these diverse forecasts yield robust, market-neutral portfolios with superior and stable out-of-sample performance. Our findings underscore the value of forecast uncertainty as a practical indicator for factor predictability, and validate the effectiveness of deep latent models combined with state-of-the-art forecasting techniques in modern asset pricing applications.

\section{Methodology}
\label{sec:methodology}

We adopt a two-stage procedure: (i) latent factor portfolios are extracted using a CAE model trained on firm-level characteristics—such as market equity, asset growth, and past return momentum—alongside asset returns; and (ii) the time series of each latent factor is forecasted using multiple models, with forecast uncertainty guiding the selection of portfolios with predictable returns.

Let \( r_{i,s} \in \mathbb{R} \) denote the excess return of asset \( i \) at time \( s \), and let \( z_{i,s-1} \in \mathbb{R}^P \) represent its lagged firm characteristics. The CAE models asset returns as:
\begin{equation}
r_{i,s} = \beta_i(z_{i,s-1})^\top f_s + u_{i,s}, 
\end{equation}
where \( \beta_i(z_{i,s-1}) \in \mathbb{R}^K \) is a nonlinear function mapping characteristics to factor loadings, \( f_s \in \mathbb{R}^K \) is a vector of latent factors shared across assets, and \( u_{i,s} \in \mathbb{R} \) is an idiosyncratic error term. The mapping \( \beta_i(\cdot) \) is parameterized as a feedforward neural network with ReLU activations:
\begin{align}
z^{(0)}_{i,s-1} &= z_{i,s-1}, \\
z^{(\ell)}_{i,s-1} &= \mathrm{ReLU}\left(W^{(\ell-1)} z^{(\ell-1)}_{i,s-1} + b^{(\ell-1)}\right), \quad \ell = 1,\dots,L_\beta, \\
\beta_i(z_{i,s-1}) &= W^{(L)} z^{(L)}_{i,s-1} + b^{(L)}.
\end{align}

The latent factors \( f_s \in \mathbb{R}^K \) are extracted via a cross-sectional projection of returns onto firm characteristics:
\begin{equation}
f_s = W_{f} \left( (Z_{s-1}^\top Z_{s-1})^{-1} Z_{s-1}^\top r_s \right), \label{eq:latent-factor}
\end{equation}
where \( Z_{s-1} \in \mathbb{R}^{N \times P} \) denotes the matrix of lagged characteristics and \( W_{f} \in \mathbb{R}^{K \times P} \) is a time-varying projection matrix. This construction aggregates firm-level information into a small number of tradable, low-noise latent portfolios. The CAE parameters \( \Theta = \{ W^{(\ell)}, b^{(\ell)}, W_{f} \} \) are estimated jointly by minimizing the cross-sectional pricing loss over the training period:
\begin{equation}
\min_{\Theta} \sum_{s=1}^t \sum_{i=1}^N \left( r_{i,s} - \beta_i(z_{i,s-1})^\top f_s \right)^2. 
\end{equation}

Once trained, the CAE produces a time series \( \{f_s^{(\kappa)}\}_{s=1}^t \) for each latent factor \( \kappa = 1, \dots, K \), which we use to generate one-step-ahead forecasts \( \hat{f}_{t+1}^{(\kappa)} \in \mathbb{R} \). In addition to point forecasts, we construct a set of quantile predictions \( \{ \hat{f}_{t+1}^{(\kappa,\alpha)} \}_{\alpha \in \mathcal{Q}} \), where \( \mathcal{Q} \subset (0,1) \) is a model-dependent collection of quantile levels. Forecast uncertainty for each factor \( \kappa \) at time \( t+1 \) is then defined as the average absolute deviation of the quantile forecasts from the central prediction:
\begin{equation}
U_{t+1,\kappa}^{(\kappa)} = \frac{1}{|\mathcal{Q}|} \sum_{\alpha \in \mathcal{Q}} \left| \hat{f}_{t+1}^{(\kappa,\alpha)} - \hat{f}_{t+1}^{(\kappa)} \right|. \label{eq:forecast-uncertainty}
\end{equation}
This general formulation accommodates all forecasting models under consideration. 

Using the forecast uncertainty estimates \( U_{t+1}^{(\kappa)} \), we rank all \( K \) latent factors in increasing order of uncertainty. For each integer \( \kappa \in \{1, \dots, K\} \), we construct a tangency portfolio using the \( \kappa \) most predictable factors. Let \( \mu_{t+1} \in \mathbb{R}^\kappa \) be the vector of forecasts for the $\kappa$ most predictable factors and \( \Sigma_f \in \mathbb{R}^{\kappa \times \kappa} \) be the covariance matrix associated with these factors computed using sample covariance matrix. The tangency portfolio is given by $w_{f,t} = \frac{\Sigma_f^{-1} \mu_{t+1}}{\mathbf{1}^\top \Sigma_f^{-1} \mu_{t+1}}$ which is then mapped to asset weights using \eqref{eq:latent-factor} via:
\begin{equation}
w_{r,t} =  Z_{t} (Z_{t}^\top Z_{t})^{-1} W_{f}^{(\kappa)\top} w_{f,t}.
\label{eq:asset-weights}
\end{equation}
where $W_{f}^{(\kappa)}\in \mathbb{R}^{\kappa \times P}$ are the columns of the CAE trained $W_{f}$ matrix that correspond to the $\kappa=1,2,\ldots,K$ selected factors. 

To implement the forecasting and uncertainty estimation defined in equations~\eqref{eq:forecast-uncertainty}–\eqref{eq:asset-weights}, we consider three models: a nonparametric sample-based baseline, a supervised regression-tree method, and a zero-shot pretrained foundational sequence model (see Table~\ref{tab:forecast_models}).

The first model, denoted IID-BS, is analogous to the original CAE approach except that it utilizes factor selection mechanism based on forecast uncertainty. In particular, it assumes that latent factor returns \( \{f_s^{(\kappa)}\}_{s=1}^t \) are drawn i.i.d. from a stationary distribution. The point forecast \( \hat{f}_{t+1}^{(\kappa)} \) is taken to be the sample mean, computed over a rolling window of size $t$ $\hat{f}_{t+1}^{(\kappa)} = \frac{1}{t} \sum_{s=1}^{t} f_s^{(\kappa)}.$

To assess predictive uncertainty, we generate \( B \) bootstrap resamples from the window and compute empirical quantiles \( \hat{f}_{t+1}^{(\kappa,\alpha)} \) at levels \( \alpha \in \{0.05, 0.95\} \). These quantiles are then used to evaluate the forecast dispersion score \( U_{t+1}^{(\kappa)} \) as defined in Equation~\eqref{eq:forecast-uncertainty}.

{\begin{table}[t]\scriptsize
\centering
\caption{Summary of Forecasting Models for Latent Factors}
\label{tab:forecast_models}
\begin{tabular}{lll}
\toprule
\textbf{Model Abbreviation} & \textbf{Forecast} & \textbf{Uncertainty Type}\\
\midrule
IID-BS & \makecell[l]{ IID-Based\\ (Sample Mean)}  & \makecell[l]{Mean Absolute Deviation from \\ the Mean across 2 Bootstrap Quantiles}\\
 & & \\
\makecell[l]{Q-Boost} & \makecell[l]{XGBoost-Based Median \\(trained with RAPIDS)} & \makecell[l]{Mean Absolute Deviation from \\ the Median across 3 Tree-Based Quantiles}\\
& & \\
\makecell[l]{ZS-Chronos} & \makecell[l]{Chronos-Based Median\\ (Zero-Shot)} & \makecell[l]{Mean Absolute Deviation from \\ the Median across 9 Chronos Quantiles} \\
\bottomrule
\end{tabular}
\end{table}}

The second model, Quantile Gradient Boosted Regression Trees (Q-Boost), estimates the conditional quantile function of each latent factor \( f_{t+1}^{(\kappa)} \) using lagged time-series features. Let \( x_t^{(\kappa)} \in \mathbb{R}^d \) denote the feature vector extracted from the history \( \{f_s^{(\kappa)}\}_{s=1}^{t} \) (see Section \ref{sec:Empirical Analysis} for details). For each quantile level \( \alpha \in \mathcal{Q} = \{0.05, 0.5, 0.95\} \), we fit a separate quantile regression tree \( Q_\kappa^{(\alpha)} \) to obtain: \\
$ \hat{f}_{t+1}^{(\kappa,\alpha)} = Q_\kappa^{(\alpha)}(x_t^{(\kappa)}). $
The median forecast $\hat{f}_{t+1}^{(\kappa,0.5)}$ serves as the point estimate, while the uncertainty score \( U_{t+1}^{(\kappa)} \) is computed from the predicted quantiles via Equation~\eqref{eq:forecast-uncertainty} using  $\hat{f}_{t+1}^{(\kappa,0.05)}$ and  $\hat{f}_{t+1}^{(\kappa,0.95)}$. We restrict the model to three quantile levels to reduce computational burden, since each \( \alpha \) requires fitting an independent learner. Despite this limitation, Q-Boost captures localized nonlinear dynamics effectively and remains computationally efficient~\cite{friedman2001greedy}.

The third model, ZS-Chronos, is a 205 million parameter pretrained Time Series Foundation Model (TSFM) based on the encoder-decoder T5 architecture \cite{ansari2024chronos}. We specifically adopt the Chronos-Bolt variant due to its GPU-optimized design and improved inference speed compared to the original Chronos model. For each factor \( \kappa \), we input its historical sequence \( \{f_s^{(\kappa)}\}_{s=1}^t \) and obtain a set of quantile forecasts \( \{ \hat{f}_{t+1}^{(\kappa,\alpha)} \}_{\alpha \in \mathcal{Q}} \) with \( \mathcal{Q} = \{0.1, 0.2, \dots, 0.9\} \). The central prediction is the median \( \hat{f}_{t+1}^{(\kappa)} = \hat{f}_{t+1}^{(\kappa,0.5)} \), and the uncertainty is computed as the mean absolute deviation from this value as in \eqref{eq:forecast-uncertainty}. Chronos is applied in a zero-shot (ZS) configuration without task-specific fine-tuning, leveraging pretrained knowledge to generate competitive forecasts across domains. Its effectiveness has been benchmarked on GIFT-Eval \cite{aksu2024gifteval}, where it ranks among the top models for financial and economic prediction tasks. This strong performance and high speed motivated our selection of Chronos as a core forecasting component. 

Together, these models provide a rich ensemble of predictive signals that support robust portfolio construction. The IID-BS model offers a model-free benchmark; Q-Boost captures local time-series structure through supervised learning; and ZS-Chronos applies deep sequence modeling in a data-efficient, pretrained framework.

\section{Empirical Analysis}\label{sec:Empirical Analysis}

We evaluate our methodology using monthly returns data for the 2000 largest US stocks, sorted by market equity at the end of each rolling estimation window. All returns are adjusted for dividends and stock splits. The predictive factors used within the CAE framework are constructed following \cite{jensen2023replication}, comprising 153 long--short portfolios spanning the US equity market from February 1962 through December 2024. For detailed information on factor construction, we refer readers to \cite{jensen2023replication} and the accompanying online appendix.

To ensure methodological rigor and eliminate potential look-ahead bias, we strictly adhere to the timing conventions detailed in \cite{jensen2023replication}. Specifically, all firm-level characteristics used in factor construction are lagged by at least six months, capturing realistic data availability constraints. This lag structure ensures that factor portfolios at time \( t \) are constructed solely from information observable no later than \( t - 6 \), preserving both causal interpretation and practical implementability within our forecasting framework.

Our empirical analysis employs an expanding-window forecasting scheme to closely replicate realistic investment conditions. Initially, the CAE is trained on data from February 1962 through December 1999 (38 years) to extract latent factor representations and characteristic-based exposures. Within each expanding training window, we utilize a rolling validation period of the most recent 12 years for hyperparameter tuning and early stopping. The out-of-sample evaluation spans January 2000 to December 2024 (25 years), during which CAE models are recalibrated annually using all historical data available up to the previous year's end, and portfolio weights are rebalanced monthly. 

For all performance analyses—including ensemble strategies—we restrict investments to the 300 largest optimal weights in absolute terms, applying normalization to exclude leverage and maintain approximate market neutrality. These practical constraints, combined with our monthly rebalancing and focus on liquid, large-cap equities, ensure that the reported strategies are robust, implementable, and resilient to market frictions.

We evaluate the performance of CAE-based models across multiple latent dimensions. Specifically, six CAE architectures are trained, each corresponding to a distinct number of latent factors, with \( K \in \{5, 10, 20, 30, 40, 50\} \). This design enables a systematic examination of model scalability and predictive efficacy as a function of latent complexity. For each value of $K$, we apply the uncertainty-aware factor selection procedure detailed in Section~\ref{sec:methodology}, which isolates the most predictable components for portfolio construction.

To reduce estimation variance and improve robustness, we ensemble multiple CAE models initialized with distinct random weight initializations, see also \cite{gu2020autoencoder} for analogous model construction and implementation details. The hyperparameters used for CAE training are summarized in Table~\ref{tab:hyperparameters_combined}.

\begin{table}[htbp]
\centering
\caption{Q-Boost Feature Set for Factor Returns Prediction}
\label{tab:features}
\resizebox{\columnwidth}{!}{
\begin{tabular}{|l|c|c|}
\hline
\textbf{Feature Name} & \textbf{Equation} & \textbf{Windows/Lags} \\
\hline
Lagged Returns & $r_{\kappa,t-\ell}$ & $\ell \in \{1,3,5,10\}$ \\
\hline
Moving Averages & $\mu^{(m)}_{\kappa,t} = \frac{1}{m} \sum_{j=0}^{m-1} r_{\kappa,t-j}$ & $m \in \{3,5,10,20\}$ \\
\hline
Rolling Std Dev & $\sigma^{(m)}_{\kappa,t} = \sqrt{\frac{1}{m-1} \sum_{j=0}^{m-1} (r_{\kappa,t-j} - \mu^{(m)}_{\kappa,t})^2}$ & $m \in \{3,5,10,20\}$ \\
\hline
Rolling Minimum & $\min^{(m)}_{\kappa,t} = \min_{j=0,...,m-1} r_{\kappa,t-j}$ & $m \in \{3,5,10,20\}$ \\
\hline
Rolling Maximum & $\max^{(m)}_{\kappa,t} = \max_{j=0,...,m-1} r_{\kappa,t-j}$ & $m \in \{3,5,10,20\}$ \\
\hline
Rate of Change & $\text{ROC}^{(\ell)}_{\kappa,t} = \frac{r_{\kappa,t}}{r_{\kappa,t-\ell}} - 1$ & $\ell \in \{1,3,5,10\}$ \\
\hline
Z-scores & $z^{(m)}_{\kappa,t} = \frac{r_{\kappa,t} - \mu^{(m)}_{\kappa,t}}{\sigma^{(m)}_{\kappa,t}}$ & $m \in \{30,60,90\}$ \\
\hline
Momentum & $\text{mom}_{\kappa,t} = \mu^{(5)}_{\kappa,t} - \mu^{(20)}_{\kappa,t}$ & 5-day vs 20-day \\
\hline
Peer Lags & $r_{\kappa',t-\ell}$ & $\ell \in \{1,3,5\}$, 3 peers \\
\hline
\end{tabular}
}
\end{table}

For latent factor forecasting, we apply three distinct techniques, IID-BS, Q-Boost, and ZS-Chronos, to capture a range of modeling assumptions and temporal dependencies. The IID-BS method generates bootstrap resamples of historical returns for each latent factor using the \texttt{numpy} library. Forecasts are formed as the sample mean, and uncertainty is quantified via empirical quantiles of the resampled distribution. This approach serves as a computationally efficient, nonparametric benchmark.

Q-Boost leverages gradient-boosted decision trees via the XGBoost framework, specifically configured for quantile regression. For each factor $\kappa$, we construct a feature vector $x_{\kappa,t} \in \mathbb{R}^d$ at time $t$ from lagged transformations of its historical return series ${r_{\kappa,s}}_{s \leq t-1}$. Our feature engineering pipeline captures a broad range of temporal patterns—including momentum effects, volatility regimes, and cross-sectional interactions with correlated peer factors. The specific mathematical formulations and parameterizations for each feature type are summarized in Table~\ref{tab:features}, resulting in $d = 37$ features per factor.

Training is accelerated using Nvidia’s GPU-based RAPIDS library \cite{rapids2018}, which provides substantial speedups over CPU execution. In contrast to Chronos’ joint sequence modeling, Q-Boost trains independent quantile regressors for each latent factor, enabling targeted learning of factor-specific dynamics.

\begin{table}[t]
\centering
\small
\caption{Model Hyperparameters for CAE and Q-Boost.}
\label{tab:hyperparameters_combined}
\begin{tabular}{|l|l|l|l|}
\hline
\multicolumn{2}{|c|}{\textbf{Panel A: CAE Model}} & \multicolumn{2}{c|}{\textbf{Panel B: Q-Boost Model}} \\
\hline
\textbf{Parameter} & \textbf{Value} & \textbf{Parameter} & \textbf{Value} \\
\hline
Learning Rate & 1e-3 & Learning Rate & 0.05 \\
Epochs & 200 & Iterations & 50 \\
Hidden Layers & [32, 16] & Depth & 3 \\
Batch Size & 10,000 & Bootstrap Type & Bayesian \\
L1 Regularization & 1e-5 & Task Type & GPU \\
Patience & 5 & Quantile Range & [0.05, 0.5, 0.95] \\
Number of Experts & 50 & Loss Function & Quantile Loss \\
Validation Period & 144 months & & \\
Retrain Frequency & 12 months & & \\
\hline
\end{tabular}
\end{table}

The Q-Boost model is trained on \( x_{\kappa,t} \) to predict the conditional quantiles of the future return \( r_{\kappa,t+1} \), using a quantile loss at levels \( \alpha \in \{0.05, 0.5, 0.95\} \), thereby yielding probabilistic forecasts that incorporate uncertainty. The specific hyperparameters used for Q-Boost are detailed in Table~\ref{tab:hyperparameters_combined}.

The ZS-Chronos model corresponds to the \texttt{Chronos-Bolt Base} architecture, a configuration of Chronos comprising 205 million parameters. We apply Chronos in a zero-shot setting using the official \texttt{chronos} library, and leveraging the \texttt{ChronosBoltPipeline} class, allowing direct inference on time-series sequences without additional training. 

\begin{figure}[b]
    \centering
    \includegraphics[width=\columnwidth]{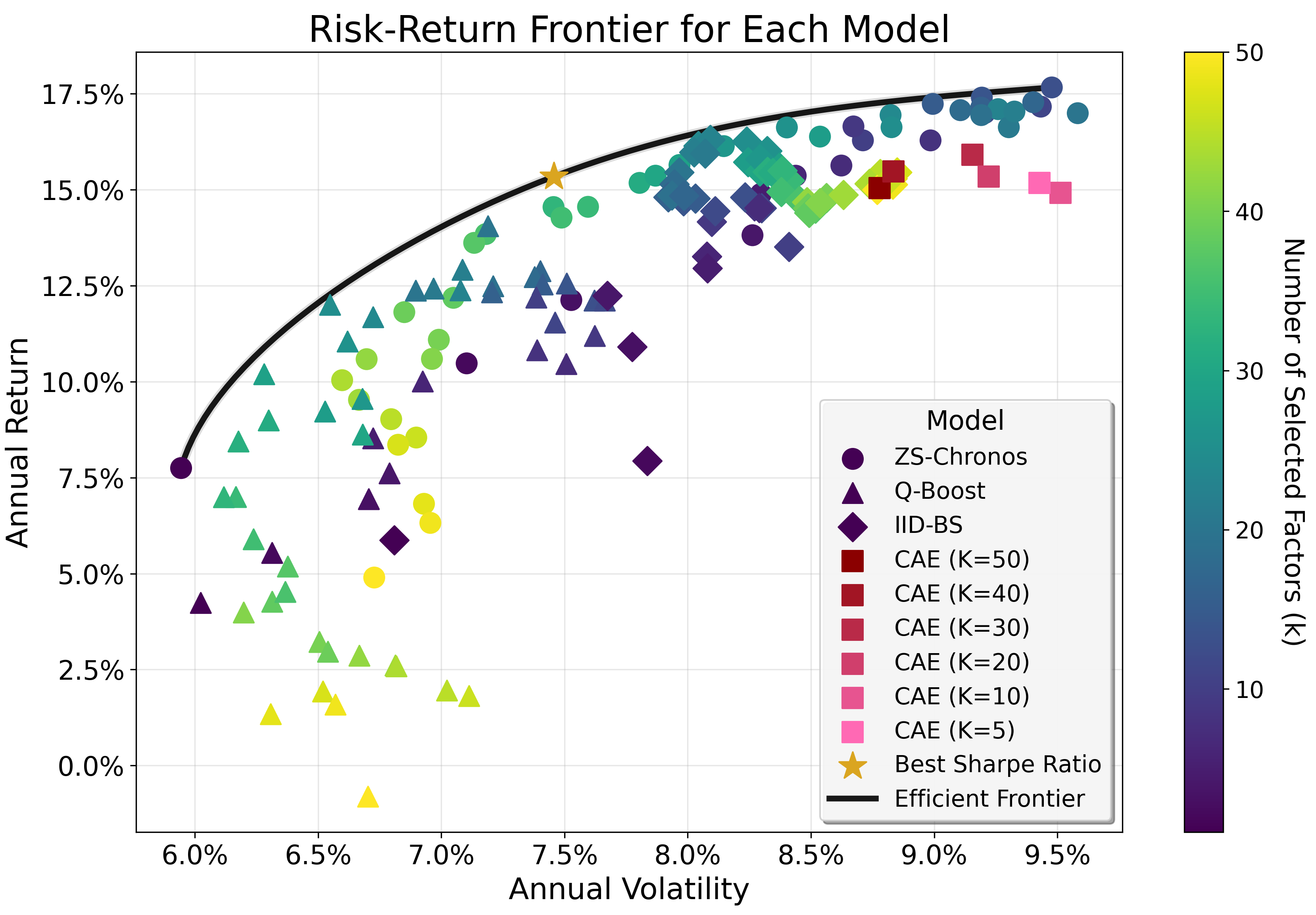}
   \caption{Out-of-sample risk–return frontier for CAE models with $K \in {5, 10, 20, 30, 40, 50}$, as well as IID-BS, Q-Boost, and ZS-Chronos (each with $K = 50$). Each point represents a model configuration constructed from a subset of $\kappa = 1, 2, \ldots, 50$ latent factors, selected according to forecast uncertainty. The figure illustrates the trade-off between annualized return and volatility across varying $\kappa$, showing that uncertainty-aware pruning consistently improves risk-adjusted performance. The smooth, concave shape of the frontier highlights model diversity, while the gradual variation in performance with the number of included factors indicates robustness to the choice of $\kappa$.}
    \label{fig:combined_frontier}
\end{figure}

We see that integrating forecast uncertainty into latent factor selection leads to substantial improvements in portfolio performance across all model architectures, as illustrated in the risk-return frontier shown in Figure~\ref{fig:combined_frontier}. This figure presents portfolios generated under various latent factor subset sizes \(\kappa\), demonstrating that models leveraging uncertainty-aware selection consistently outperform the baseline CAE models. The plotted configurations reflect all combinations generated under the K=50 setting for the IID-BS, Q-Boost, and ZS-Chronos models, as well as the configurations from the original CAE benchmarks.

Our analysis illustrates a smooth curvature in the frontier plot as \(\kappa\) changes, reflecting a stable relationship between the number of selected latent factors and portfolio performance. This behavior suggests that the procedure is not highly sensitive to small perturbations in \(\kappa\), thereby reducing the risk of overfitting. The observed performance improvements are therefore attributable to the exploitation of persistent structure in the data rather than spurious variation.

Our findings indicate that strong out-of-sample performance can be achieved without utilizing the full set of latent factors. Configurations with \(\kappa < 50\) often outperform the full CAE benchmark set, with \(\kappa \in \{20, \ldots, 40\}\) frequently offering the best return-volatility trade-off. Our results, detailed in Table~\ref{tab:main_results}, demonstrate that selectively choosing predictable factors enhances portfolio efficiency by concentrating exposure on stable drivers of return while reducing exposure to high-variance components.

\begin{table*}[!t]
\centering
\caption{Performance comparison of each model (CAE, IID-BS, Q-Boost, ZS-Chronos) and ensemble strategies over 2000--2024. Each row corresponds to a specific latent dimensionality $K$ and its best-performing factor subset size $\kappa^*$. Metrics include Sortino, Sharpe, Omega ratios, annualized return, volatility, and maximum drawdown. Adaptive $\kappa$ models adjust factor size based on past performance. Ensemble~(A) includes SPY and achieves the best risk-adjusted metrics; Ensemble~(B) excludes SPY and yields the highest total and annualized returns. Bold and underlined values indicate best overall performance.}
\scriptsize
\begin{tabular}{lcccccccc}
\toprule
\textbf{Model} & \textbf{$K$} & \textbf{$\kappa^*$} & \textbf{Sortino Ratio} & \textbf{Sharpe Ratio} & \textbf{Omega Ratio} & \textbf{Ann. Ret. (\%)} & \textbf{Ann. Vol. (\%)} & \textbf{Max DD. (\%)} \\

\midrule
CAE & 5 & -- & \textbf{3.053} & 1.610 & 3.437 & 15.18 & 9.43 & 16.57 \\
CAE & 10 & -- & 2.729 & 1.568 & 3.382 & 14.92 & 9.51 & \textbf{10.48} \\
CAE & 20 & -- & 2.780 & 1.664 & 3.596 & 15.35 & 9.22 & 13.36 \\
CAE & 30 & -- & 2.949 & 1.738 & 3.814 & \textbf{15.91} & 9.15 & 19.48 \\
CAE & 40 & -- & 2.908 & \textbf{1.751} & \textbf{3.922} & 15.47 & \textbf{8.84} & 18.29 \\
CAE & 50 & -- & 2.866 & 1.714 & 3.803 & 15.05 & 8.78 & 18.07 \\

\midrule
IID-BS & 5 & 5 & 3.042 & 1.604 & 3.419 & 15.15 & 9.45 & 16.57 \\
IID-BS & 10 & 4 & 3.239 & 1.642 & 3.758 & 15.02 & 9.15 & 10.65 \\
IID-BS & 20 & 3 & 3.262 & 1.647 & 3.969 & 12.54 & \textbf{7.62} & \textbf{8.94} \\
IID-BS & 30 & 27 & 3.111 & 1.827 & 4.092 & \underline{\textbf{16.77}} & 9.18 & 15.07 \\
IID-BS & 40 & 16 & 3.424 & 1.854 & 4.261 & 15.55 & 8.39 & 15.95 \\
IID-BS & 50 & 25 & \textbf{3.714} & \textbf{1.973} & \textbf{4.841} & 16.26 & 8.24 & 11.76 \\
IID-BS & 50 & Adaptive & 3.565 & 1.924 & 4.640 & 16.03 & 8.33 & 12.20 \\

\midrule
Q-Boost & 5 & 3 & 2.520 & 1.520 & 3.168 & 14.14 & 9.31 & 15.16 \\
Q-Boost & 10 & 5 & 2.485 & 1.508 & 3.296 & \textbf{14.46} & 9.59 & 14.19 \\
Q-Boost & 20 & 4 & 3.073 & 1.603 & 3.574 & 11.50 & 7.18 & \underline{\textbf{7.54}} \\
Q-Boost & 30 & 10 & 2.721 & 1.694 & 3.658 & 12.10 & 7.14 & 10.90 \\
Q-Boost & 40 & 17 & 2.806 & 1.599 & 3.376 & 10.68 & \textbf{6.68} & 11.52 \\
Q-Boost & 50 & 20 & \textbf{3.351} & \textbf{1.954} & \textbf{4.696} & 14.05 & 7.19 & 10.57 \\
Q-Boost & 50 & Adaptive & 3.080 & 1.700 & 3.812 & 12.80 & 7.53 & 18.92 \\
  
\midrule
ZS-Chronos & 5 & 5 & 2.840 & 1.663 & 3.505 & 15.93 & 9.58 & 18.10 \\
ZS-Chronos & 10 & 6 & 2.876 & 1.703 & 3.629 & 15.79 & 9.27 & 13.59 \\
ZS-Chronos & 20 & 8 & 3.379 & 1.762 & 3.737 & 13.81 & 7.84 & 13.38 \\
ZS-Chronos & 30 & 15 & 3.311 & 1.808 & 4.156 & 13.73 & \textbf{7.59} & \textbf{8.49} \\
ZS-Chronos & 40 & 8 & \textbf{3.984} & \textbf{1.905} & 4.901 & 16.29 & 8.55 & 11.17 \\
ZS-Chronos & 50 & 7 & 3.847 & 1.813 & 4.608 & 15.63 & 8.62 & 9.25 \\
ZS-Chronos & 50 & Adaptive & 3.042 & 1.872 & \textbf{5.169} & \textbf{16.62} & 8.88 & 15.62 \\

\midrule
Ensemble (A) & 50 & Adaptive & 4.010 & \underline{\textbf{2.204}} & \underline{\textbf{5.952}} & 14.37 & \underline{\textbf{6.52}} & \textbf{9.22} \\
Ensemble (B) & 50 & Adaptive & \underline{\textbf{4.222}} & 2.111 & 5.433 & \textbf{15.05} & 7.13 & 12.13 \\

\bottomrule
\end{tabular}
\label{tab:main_results}
\end{table*}
\normalsize

While the preceding analysis demonstrates that post-hoc selection of $\kappa$ yields robust performance, it implicitly assumes future optimality of $\kappa$, limiting practical applicability. To address this limitation, we introduce an adaptive selection procedure for the number of latent factors, denoted by $\kappa_t^{(m)*}$, relying solely on information available at decision point $t$. This adaptation preserves the uncertainty-aware factor selection methodology while offering realistic and implementable forecasts.

Specifically, we employ an expanding-window framework to dynamically determine $\kappa_t^{(m)*}$ for each predictive model $m \in \{1,2,3\}$ (IID-BS, Q-Boost, ZS-Chronos). Instead of using a fixed $\kappa$, we adaptively optimize a temporally regularized objective based on past risk-adjusted performance. We opt for the Sortino ratio, defined below, as our main selection criterion because it emphasizes downside risk and thus facilitates a more aggressive adaptive choice of $\kappa$ compared to the Sharpe ratio, which penalizes both upside and downside deviations equally. While Sharpe could be an alternative, our objective is to prioritize portfolios less susceptible to negative returns.

Formally, let $r_t^{(m,\kappa)} \in \mathbb{R}^H$ denote the out-of-sample return vector for model $m$ with $\kappa$ factors over the last $H$ periods. Define empirical downside deviation, mean return, and the Sortino ratio as:
\begin{align*}
\sigma^{-}_t{}^{(m,\kappa)} &= \sqrt{ \frac{1}{H}\sum_{h=1}^{H}\left(\min\{r_{t,h}^{(m,\kappa)},0\}\right)^2 },\\
\mu_t^{(m,\kappa)} &= \frac{1}{H}\sum_{h=1}^{H}r_{t,h}^{(m,\kappa)}, \quad \mathrm{SoR}_t^{(m)}(\kappa) = \frac{\mu_t^{(m,\kappa)}}{\sigma^{-}_t{}^{(m,\kappa)}+\varepsilon}, \quad \varepsilon>0.
\end{align*}

Direct maximization of the Sortino ratio over $\kappa$ at each step is unstable due to high variance induced by occasional negative returns. Thus, we propose a smooth, differentiable approximation of the discrete optimization via a log-sum-exp (LSE) formulation. Introducing the latent variable $\theta_t^{(m)} = \log\kappa_t^{(m)*}$ (log-scale for positivity), the objective becomes:
\[
\mathcal{L}_{\mathrm{LSE}}^{(m)}(\theta_t^{(m)})=\frac{1}{\lambda}\log\sum_{\kappa\in K_m}\exp\left(\lambda\cdot\mathrm{SoR}_t^{(m)}(\kappa)-\lambda(\theta_t^{(m)}-\log\kappa)^2\right),
\]
where $\lambda>0$ controls smoothness, and the quadratic term penalizes deviations in log-space, providing a localizing effect that stabilizes the selection.

To enhance temporal stability, we include a regularization term encouraging gradual evolution of factor complexity over time: $\mathcal{R}_{\mathrm{smooth}}^{(m)}(\theta_t^{(m)})=\frac{\eta}{2}\left(\theta_t^{(m)}-\theta_{t-1}^{(m)}\right)^2$ for $\eta>0$. The final optimization solved at each window is\\ $\theta_t^{(m)*} = \arg\min_{\theta\in\mathbb{R}}\left\{-\mathcal{L}_{\mathrm{LSE}}^{(m)}(\theta)+\mathcal{R}_{\mathrm{smooth}}^{(m)}(\theta)\right\}$, 
from which we obtain $\kappa_t^{(m)*}=\mathrm{round}\left(\exp(\theta_t^{(m)*})\right)$.
\begin{figure}[t]
    \centering
    \includegraphics[width=\columnwidth]{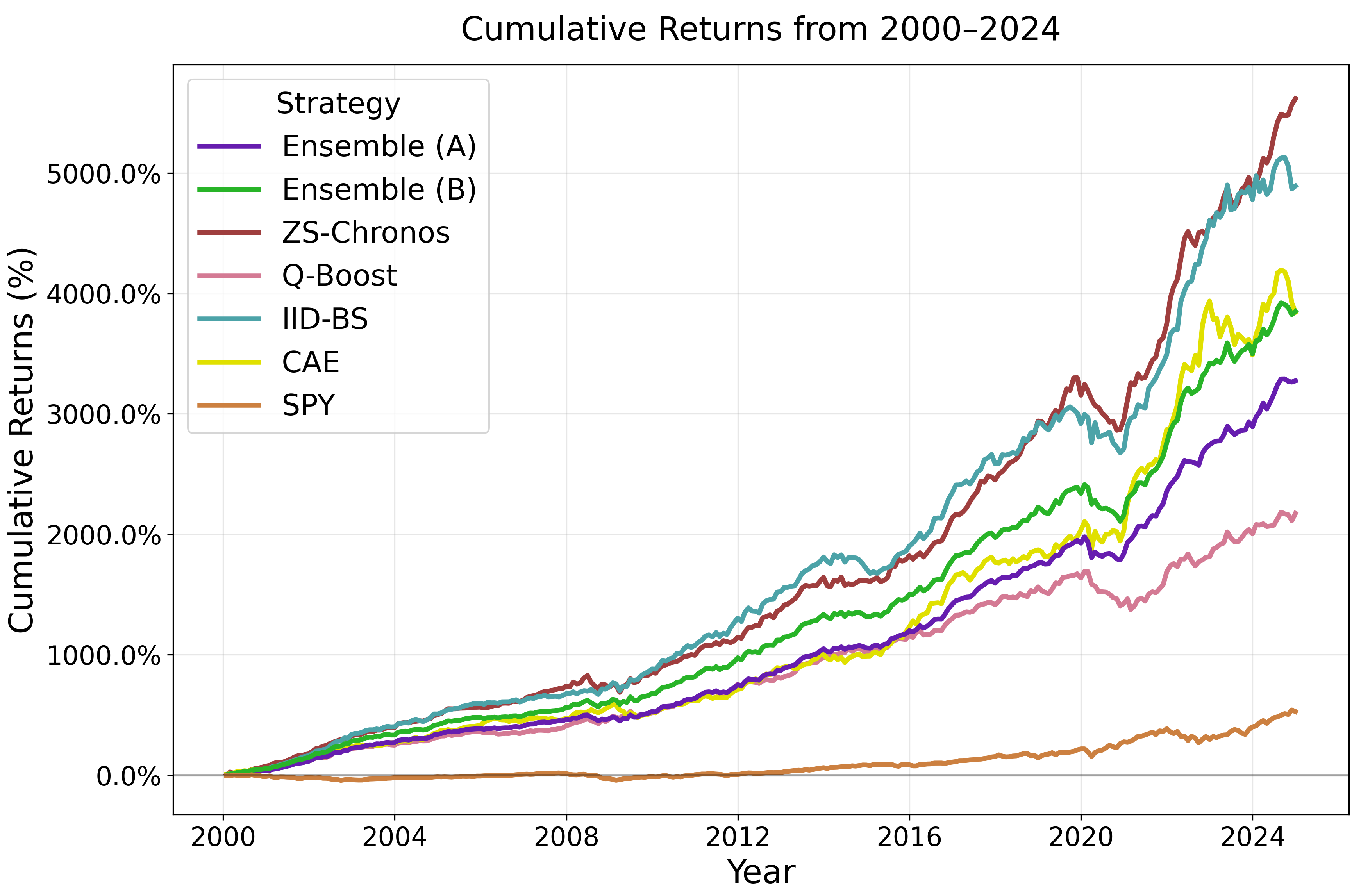}
    \caption{Cumulative returns from 2000 to 2024 for each forecasting model using adaptive $\kappa^*$ selection. The plot compares individual strategies—ZS-Chronos, Q-Boost, IID-BS, CAE (with $K=5$)—against the SPY benchmark and two ensemble portfolios. Ensemble~(A) includes SPY and achieves the highest Sharpe and Sortino ratios, while Ensemble~(B), constructed from adaptive strategies only, yields the highest total return and annualized growth. All models significantly outperform SPY, demonstrating the effectiveness of uncertainty-aware latent factor selection.}
    \label{fig:cumulative_returns}
\end{figure}

\begin{figure}[t]
    \centering
    \includegraphics[width=\columnwidth]{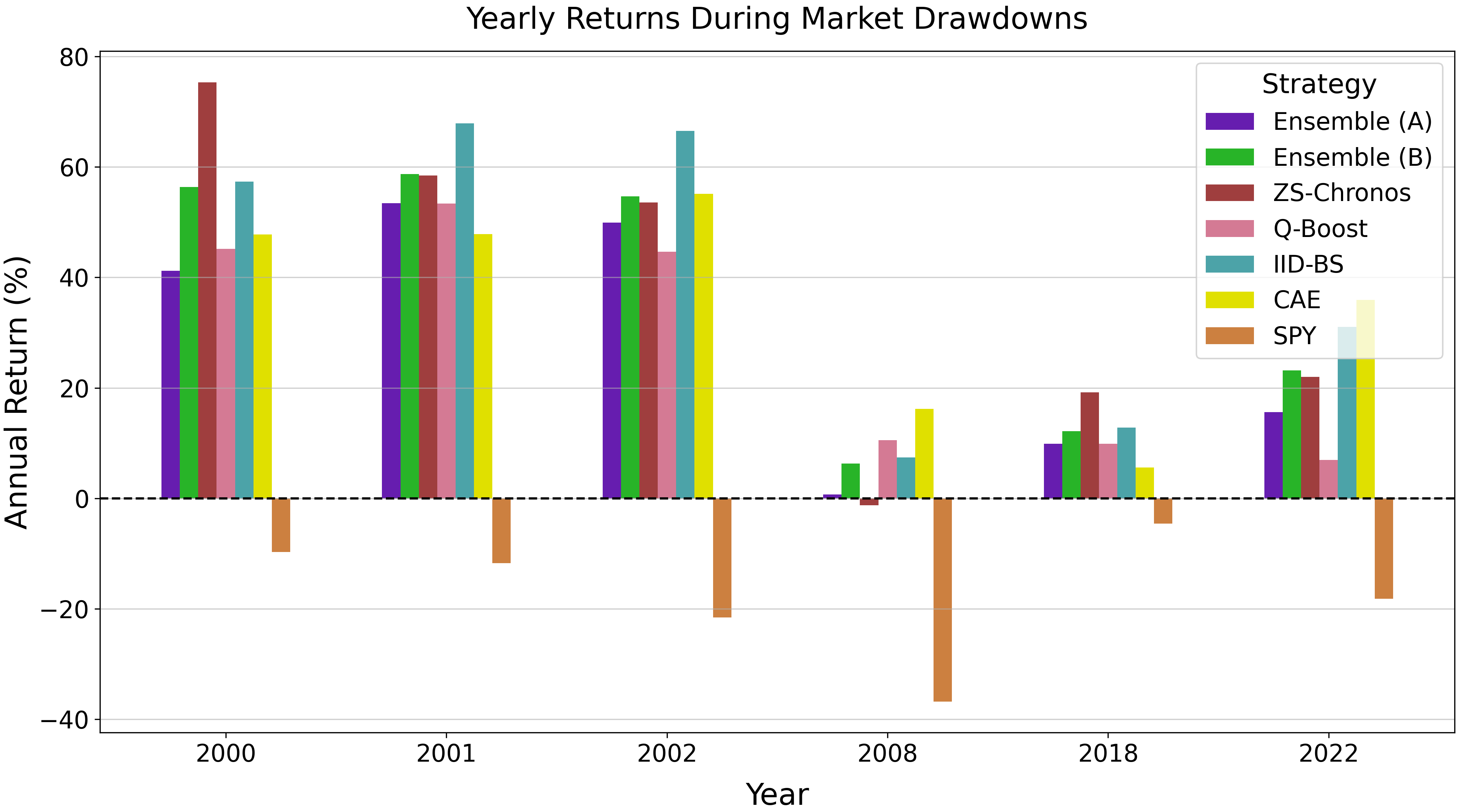}
    \caption{Yearly returns of each (adaptive $\kappa^*$) strategy during market drawdown years—defined as years when the benchmark index (SPY) posted negative annual returns. All strategies achieve positive performance during these periods, with the sole exception of ZS-Chronos, which incurred a modest loss of 1.25\% in 2008.}
    \label{fig:eoy_returns}
\end{figure}

This log-sum-exp approximation follows the smooth optimization framework of \cite{Nesterov2005Smooth}, enabling stable gradient-based optimization. Additionally, our regularization term is conceptually aligned with the Follow-The-Regularized-Leader (FTRL) method \cite{McMahan2017FTRL,Hazan2017Survey}, which imposes stability through proximity to previous solutions. In empirical analyses, we set $\lambda=1$, $\eta=2$, and the lookback period $H=12$, initializing $\kappa_t^{(m)*}$ as $K/2$ for the first 12 periods. We empirically evaluated the proposed adaptive selection method and found the results to be robust across a range of parameter choices ($\lambda$, $\eta$, and $H$), confirming the practical effectiveness and generalizability of our framework.

We observe that the individual adaptive strategies, IID-BS, Q-Boost, and ZS-Chronos, demonstrate consistently strong performance over the full out-of-sample period from 2000 to 2024. As shown in Figure~\ref{fig:cumulative_returns}, each strategy outperforms the market (SPY) with higher cumulative returns. Their resilience is evident during market downturn periods, Figure~\ref{fig:eoy_returns} highlights that almost all models produce positive returns even in years with market drawdowns. 

\begin{figure}[t]
    \centering
    \includegraphics[width=0.9\columnwidth]{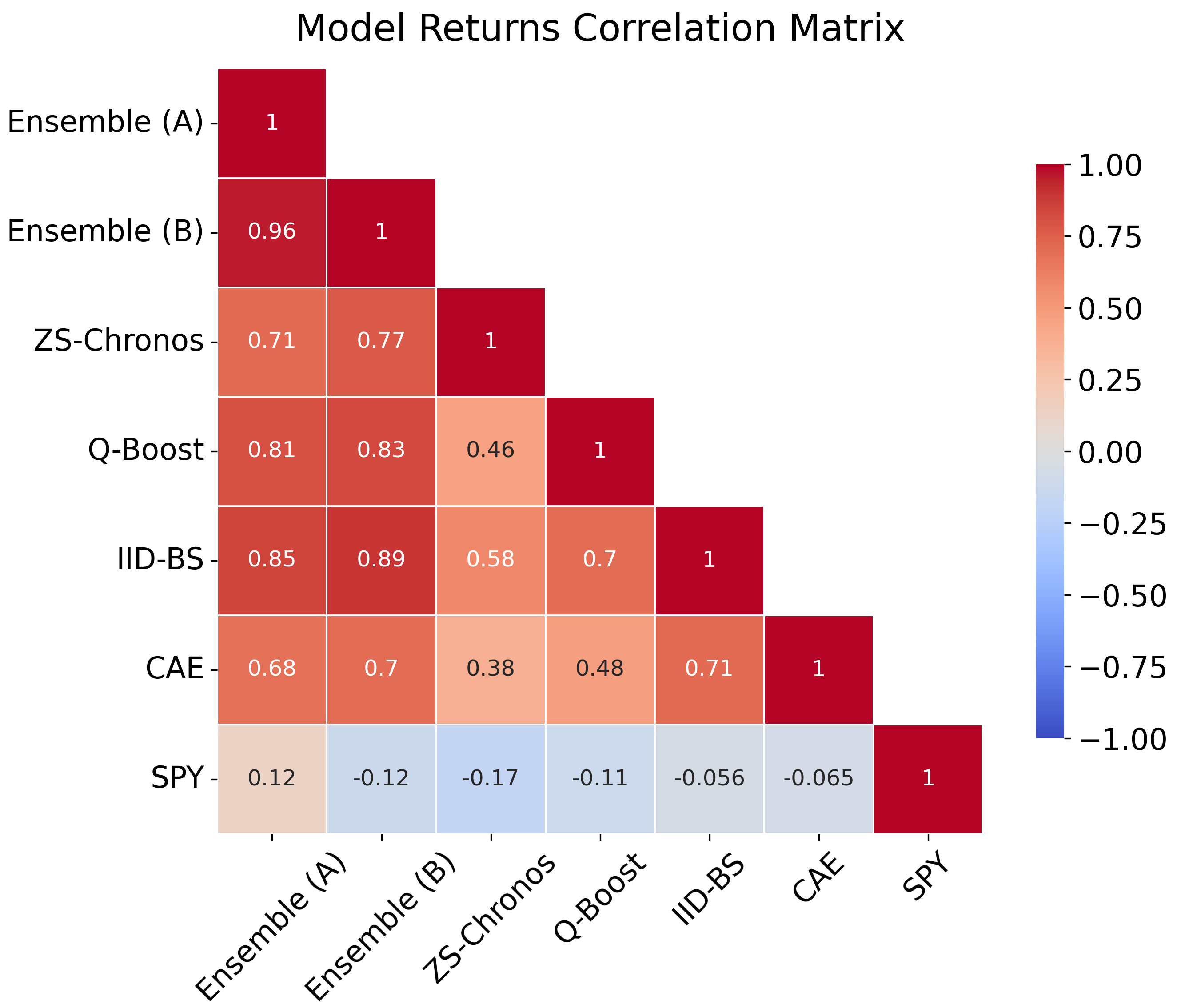}
    \caption{Correlation matrix of out-of-sample returns from 2000 to 2024 across adaptive $\kappa^*$ strategies. Each entry represents the Pearson correlation between two models’ return series, including ZS-Chronos, Q-Boost, IID-BS, CAE ($K=5$), SPY, and the two ensemble portfolios. The low correlations among strategies highlight their complementary predictive signals and justify ensemble construction.}
    \label{fig:correlation_matrix}
\end{figure}

Despite sharing the same CAE latent factor universe ($K=50$), the adaptive strategies exhibit low pairwise return correlations, as illustrated in Figure~\ref{fig:correlation_matrix}. This diversity shows that each predictive model captures distinct signals and reacts differently to underlying factor dynamics and market conditions.

\begin{table*}[t]
\centering
\resizebox{0.8\textwidth}{!}{
  \begin{tabular}{@{}l c c c c c c c@{}}
    \hline
    \multicolumn{8}{c}{\textbf{Panel A: Out-of-Sample Performance Metrics}} \\
    \hline
     & Ensemble (A) & Ensemble (B) & ZS‑Chronos & Q‑Boost & IID‑BS & CAE ($K=5$) & SPY \\
    \hline
    Total Return (\%)      & 3274.15 & 3849.92 & \underline{\textbf{5615.46}} & 2173.80 & 4890.61 & 3842.66 & 528.38 \\
    CAGR (\%)              & 15.114 & 15.841 & \underline{\textbf{17.566}} & 13.310 & 16.930 & 15.833 & 7.629 \\
    Sharpe Ratio           & \underline{\textbf{2.204}} & 2.111 & 1.876 & 1.704 & 1.928 & 1.613 & 0.560 \\
    Sortino Ratio          & 4.010 & \underline{\textbf{4.222}} & 3.052 & 3.090 & 3.577 & 3.064 & 0.791 \\
    Omega Ratio            & \underline{\textbf{5.952}} & 5.433 & 5.169 & 3.812 & 4.640 & 3.437 & 1.513 \\
    Annual Return (\%)     & 14.367 & 15.045 & \underline{\textbf{16.675}} & 12.842 & 16.085 & 15.226 & 8.551 \\
    Annual Volatility (\%) & \underline{\textbf{6.520}} & 7.127 & 8.886 & 7.538 & 8.343 & 9.440 & 15.265 \\
    Max Drawdown (\%)      & \underline{\textbf{9.224}} & 12.126 & 15.067 & 17.681 & 12.057 & 15.789 & 50.785 \\
    Market (SPY) Beta      & 0.051 & -0.055 & -0.100 & -0.055 & \underline{\textbf{-0.031}} & -0.040 & 1 \\
    Market (SPY) Alpha (\%) & 13.934 & 15.517 & \underline{\textbf{17.532}} & 13.313 & 16.348 & 15.571 & 0 \\
    \multicolumn{8}{c}{} \\
    \hline
    \multicolumn{8}{c}{\textbf{Panel B: Performance with 10 Basis Points Transaction Costs}} \\
    \hline
    Total Return (\%)      & 1773.13 & 2018.46 & \underline{\textbf{2974.48}} & 1124.51 & 2501.68 & 1987.92 & -- \\
    CAGR (\%)              & 12.44 & 12.94 & \underline{\textbf{14.63}} & 10.50 & 13.87 & 12.88 & -- \\
    Sharpe Ratio           & \underline{\textbf{1.830}} & 1.770 & 1.596 & 1.378 & 1.620 & 1.341 & -- \\
    Sortino Ratio          & 3.208 &\underline{\textbf{3.478}} & 2.637 & 2.445 & 2.961 & 2.473 & -- \\
    Omega Ratio            & \underline{\textbf{4.344}} & 4.106 & 4.050 & 2.959 & 3.607 & 2.800 & -- \\
    Annual Return (\%)     & 11.99 & 12.48 & \underline{\textbf{14.11}} & 10.31 & 13.40 & 12.61 & -- \\
    Annual Volatility (\%)  & \underline{\textbf{6.55}} & 7.05 & 8.84 & 7.48 & 8.27 & 9.40 & -- \\
    Max Drawdown (\%)      & \underline{\textbf{12.35}} & 14.33 & 17.41 & 21.60 & 15.76 & 18.04 & -- \\
    Monthly Turnover       & \underline{\textbf{1.961}} & 2.092 & 2.047 & 2.080 & 2.094 & 2.110 & -- \\
    \multicolumn{8}{c}{} \\
    \hline
    \multicolumn{8}{c}{\textbf{Panel C: Monthly Alphas from Expanding Factor Analysis}} \\
    \hline
    Monthly Alpha (\%)     & 1.016*** & 1.078*** & \underline{\textbf{1.237}}*** & 0.919*** & 1.188*** & 1.106*** & -- \\
                           & (9.53) & (9.21) & (8.42) & (7.35) & (8.56) & (7.56) & \\
    + Mkt-RF               & 0.988*** & 1.117*** & \underline{\textbf{1.287}}*** & 0.956*** & 1.204*** & 1.153*** & -- \\
                           & (9.23) & (9.55) & (8.76) & (7.63) & (8.59) & (7.87) & \\
    + SMB                  & 0.986*** & 1.113*** & \underline{\textbf{1.264}}*** & 0.964*** & 1.198*** & 1.153*** & -- \\
                           & (9.20) & (9.51) & (8.96) & (7.72) & (8.56) & (7.86) & \\
    + HML                  & 0.972*** & 1.101*** & \underline{\textbf{1.262}}*** & 0.952*** & 1.183*** & 1.128*** & -- \\
                           & (9.14) & (9.46) & (8.92) & (7.66) & (8.51) & (7.85) & \\
    + RMW                  & 0.896*** & 1.043*** & \underline{\textbf{1.271}}*** & 0.879*** & 1.119*** & 1.016*** & -- \\
                           & (8.28) & (8.75) & (8.71) & (6.92) & (7.84) & (6.97) & \\
    + CMA                  & 0.839*** & 0.985*** & \underline{\textbf{1.189}}*** & 0.849*** & 1.062*** & 0.960*** & -- \\
                           & (7.82) & (8.33) & (8.27) & (6.64) & (7.45) & (6.59) & \\
    + MOM                  & 0.811*** & 0.952*** & \underline{\textbf{1.148}}*** & 0.826*** & 1.038*** & 0.933*** & -- \\
                           & (7.84) & (8.25) & (8.18) & (6.50) & (7.33) & (6.46) & \\
    + STR                  & 0.791*** & 0.939*** & \underline{\textbf{1.154}}*** & 0.809*** & 1.012*** & 0.903*** & -- \\
                           & (7.87) & (8.39) & (8.24) & (6.63) & (7.79) & (6.95) & \\
    Final $R^2$            & 0.222 & 0.197 & 0.202 & 0.164 & 0.230 & \underline{\textbf{0.307}} & -- \\
    \hline
  \end{tabular}
}
\caption{Out-of-sample performance metrics and expanding factor regressions from January 2000 to December 2024 for portfolios constructed using adaptive $\kappa^*$ selection. 
Panel~A reports performance statistics, including total and annualized returns, volatility, Sharpe, Sortino, and Omega ratios, as well as CAPM alpha and beta relative to the SPY benchmark. 
Panel~B reports performance metrics with 10 basis points transaction costs, where monthly turnover for each strategy is averaged across the out-of-sample backtest period.
Panel~C presents monthly alphas from time-series regressions of excess portfolio returns on an expanding sequence of Fama–French, momentum and short-term reversal factors, computed on gross returns. 
Each row adds one additional factor to the regression, and the reported alpha represents the intercept after controlling for that factor and all preceding factors in the sequence. 
The ordering of factors is Market (Mkt–RF), Size (SMB), Value (HML), Profitability (RMW), Investment (CMA), Momentum (MOM), and Short-Term Reversal (STR). 
$t$-statistics are reported in parentheses. Significance levels: *~$p<0.10$, **~$p<0.05$, ***~$p<0.01$.}
\label{tab:performance_metrics_adaptive}
\end{table*}

Driven by the complementary performance of the individual strategies, we construct two ensemble portfolios using a tangency portfolio framework. At each rebalancing date \( t \), we compute the sample mean return vector \( \mu_t \in \mathbb{R}^n \) and the sample covariance matrix \( \Sigma_t \in \mathbb{R}^{n \times n} \), where \( n \) denotes the number of constituent strategies. These estimates are derived using an expanding window of all available out-of-sample returns observed up to time \( t \). The portfolio weights are then determined by maximizing the Sharpe ratio, giving the tangency portfolio: \( w_t = \frac{\Sigma_t^{-1} \mu_t}{\mathbf{1}^\top \Sigma_t^{-1} \mu_t} \), where \( w_t \in \mathbb{R}^n \) is the vector of optimal portfolio weights among each strategy.

We consider two variants of this ensemble construction. Ensemble (A) includes the market index (SPY) alongside the adaptive strategies, while Ensemble (B) allocates exclusively among the adaptive models, thereby maintaining full market neutrality. As reported in Table~\ref{tab:performance_metrics_adaptive}, both ensemble strategies exhibit superior performance across a broad range of evaluation metrics. Ensemble (A) achieves the highest Sharpe ratio (2.20), and Omega ratio (5.95), as well as the lowest annualized volatility (6.52\%) and lowest maximum drawdown (9.22\%). Ensemble (B) achieves the highest Sortino ratio (4.22), while maintaining negative market beta.

To evaluate the performance of our model accounting for transaction costs, we compute the monthly turnover as the $\ell_1$ norm $TO_t = \sum_{i=1}^N |w^\dagger_{(t-1),i} - w_{t,i}| = \|\mathbf{w}_t^\dagger - \mathbf{w}\|_1,$ where $\mathbf{w}_t^\dagger$ denotes the portfolio weights immediately before rebalancing at time $t$, carried forward from the previous allocation after returns have been realized. The net returns are computed using a linear transaction cost model:
$ 
R^{Net}_{p,t} = \left(1 - \kappa \ TO_t\right)\left(1 + R^{Gross}_{p,t}\right) - 1, 
$
where $R^{Gross}_{p,t}$ is the portfolio's gross return before transaction costs and $\kappa>0$ is the transaction cost proportionality constant, e.g., $\kappa =0.001$ corresponds to $10$ bps costs. The investment universe consists of large-cap and mid-cap stocks within the top 2,000 by market equity, updated each month on a rolling basis. This ensures high liquidity and supports realistic transaction costs with 10 bps.

Given the portfolio's gross exposure constraint of 2.0, the theoretical maximum monthly turnover is 4.0, corresponding to complete liquidation and reconstruction of both long and short positions. The observed average monthly turnover of approximately 2.0 indicates that models rebalance roughly half their positions each month, demonstrating moderate trading intensity relative to the maximum possible. Table~\ref{tab:performance_metrics_adaptive} Panel~B summarizes the average monthly turnover, and performance metrics for each model after accounting for transaction costs.

\subsection{Ablation Study of Chronos Look-Ahead Bias}

Given the strong empirical performance of the ZS-Chronos model across the evaluation period, it is natural to ask whether this success might reflect any hidden biases introduced during pretraining, particularly from exposure to financial data. This concern is especially relevant for foundation models trained on broad, heterogeneous datasets where information leakage is often difficult to trace. To address this, we conduct an ablation study analyzing both the origin of the input data used in our evaluation and the detailed characteristics of the Chronos model’s pretraining data.

Notably, the inputs provided to ZS-Chronos in our framework are not raw market time series, but rather synthetic latent factor return sequences generated by the CAE model. These sequences are constructed as weighted linear combinations of stock returns using CAE-learned factor loadings, making them unique to our experimental setup. Because these latent factors are derived internally and did not exist prior to model training, there is no possibility that ZS-Chronos encountered them during pretraining. 

Additionally, we analyze the specific financial data used during Chronos pretraining. As documented in \cite{ansari2024chronos}, the only financial dataset included in the training corpus was the M4 dataset, developed for the M4 Forecasting Competition. This dataset comprises 100,000 historical time series extracted from the ForeDeCk database and was finalized on December 28, 2017 \cite{Makridakis2020M4, Makridakis2018M4}. Consequently, no financial data beyond January 1, 2018 was available to the Chronos model during training, ensuring that its forecasts for our evaluation period are not influenced by any post-2017 market information.

To evaluate this, we compare the performance of ZS-Chronos and the benchmark models across two separate periods: before M4 dataset availability (2011–2017) and after (2018–2024). These two seven-year windows provide a balanced view of performance across distinct market regimes and enable us to examine whether ZS-Chronos model demonstrates degradation after the training cutoff point. If pretraining on the M4 dataset introduced any forward-looking bias, we would expect a measurable decline in performance in the post-2018 period.

\begin{table}[t]
\centering
\caption{Performance Statistics Before (2011--2017) and After (2018--2024)}
\label{tab:chronos_combined}
\resizebox{\columnwidth}{!}{
\begin{tabular}{llcccc}
\toprule
\textbf{Period} & \textbf{Metric} & \textbf{ZS-Chronos} & \textbf{Q-Boost} & \textbf{IID-BS} & \textbf{CAE (K=5)} \\
\midrule
\multirow{9}{*}{\textbf{Before}} 
& Total Return (\%)      & 132.77 & 108.01 & 128.16 & \underline{\textbf{159.27}} \\
& CAGR (\%)              & 12.83 & 11.03 & 12.51 & \underline{\textbf{14.58}} \\
& Annual Return (\%)     & 12.31 & 10.65 & 12.05 & \underline{\textbf{13.95}} \\
& Annual Volatility (\%) & 6.06 & \underline{\textbf{5.36}} & 6.49 & 7.30 \\
& Sharpe Ratio           & \underline{\textbf{2.033}} & 1.989 & 1.856 & 1.912 \\
& Sortino Ratio          & 3.381 & 3.591 & 3.815 & \underline{\textbf{3.900}} \\
& Omega Ratio            & \underline{\textbf{4.481}} & 4.258 & 3.760 & 3.986 \\
& Max Drawdown (\%)      & 4.44 & \underline{\textbf{3.22}} & 7.99 & 5.92 \\
& Market (SPY) Beta      & 0.05 & \underline{\textbf{0.02}} & 0.10 & \underline{\textbf{0.02}} \\
\midrule
\multirow{9}{*}{\textbf{After}} 
& Total Return (\%)      & \underline{\textbf{124.10}} & 50.17 & 85.81 & 111.19 \\
& CAGR (\%)              & \underline{\textbf{12.22}} & 5.98 & 9.25 & 11.27 \\
& Annual Return (\%)     & \underline{\textbf{11.79}} & 6.10 & 9.21 & 11.28 \\
& Annual Volatility (\%) & \underline{\textbf{6.46}} & 7.44 & 8.08 & 10.54 \\
& Sharpe Ratio           & \underline{\textbf{1.826}} & 0.820 & 1.140 & 1.070 \\
& Sortino Ratio          & \underline{\textbf{3.898}} & 1.258 & 1.717 & 1.960 \\
& Omega Ratio            & \underline{\textbf{3.665}} & 1.846 & 2.410 & 2.260 \\
& Max Drawdown (\%)      & 12.83 & 17.68 & 12.06 & \underline{\textbf{11.09}} \\
& Market (SPY) Beta      & -0.07 & -0.07 & \underline{\textbf{-0.03}} & 0.04 \\
\bottomrule
\end{tabular}
}
\end{table}

\begin{figure}[t]
    \centering
    \includegraphics[width=0.9\columnwidth]{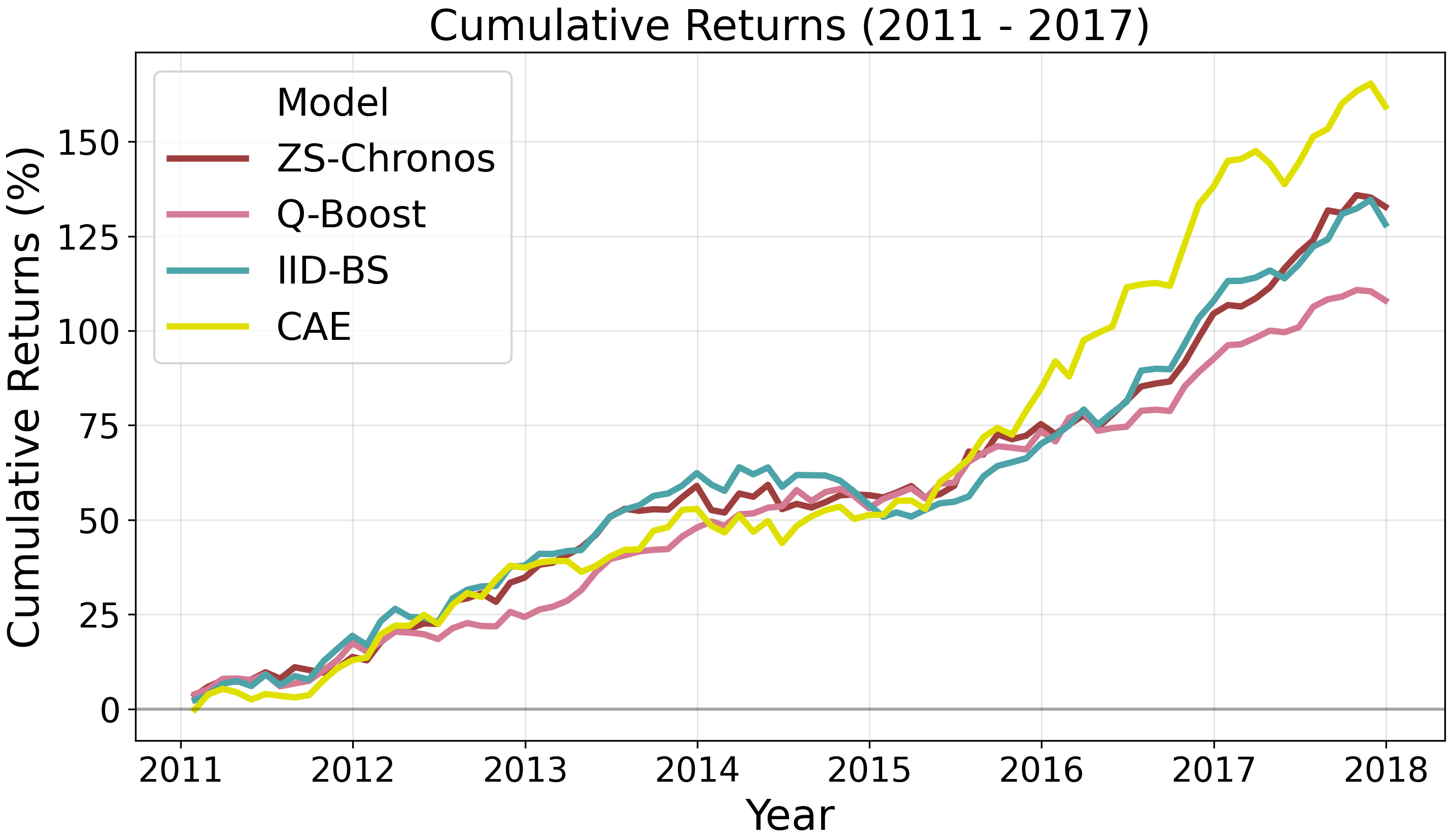}
    \caption*{(a) Model performance before dataset creation.}
    \includegraphics[width=0.9\columnwidth]{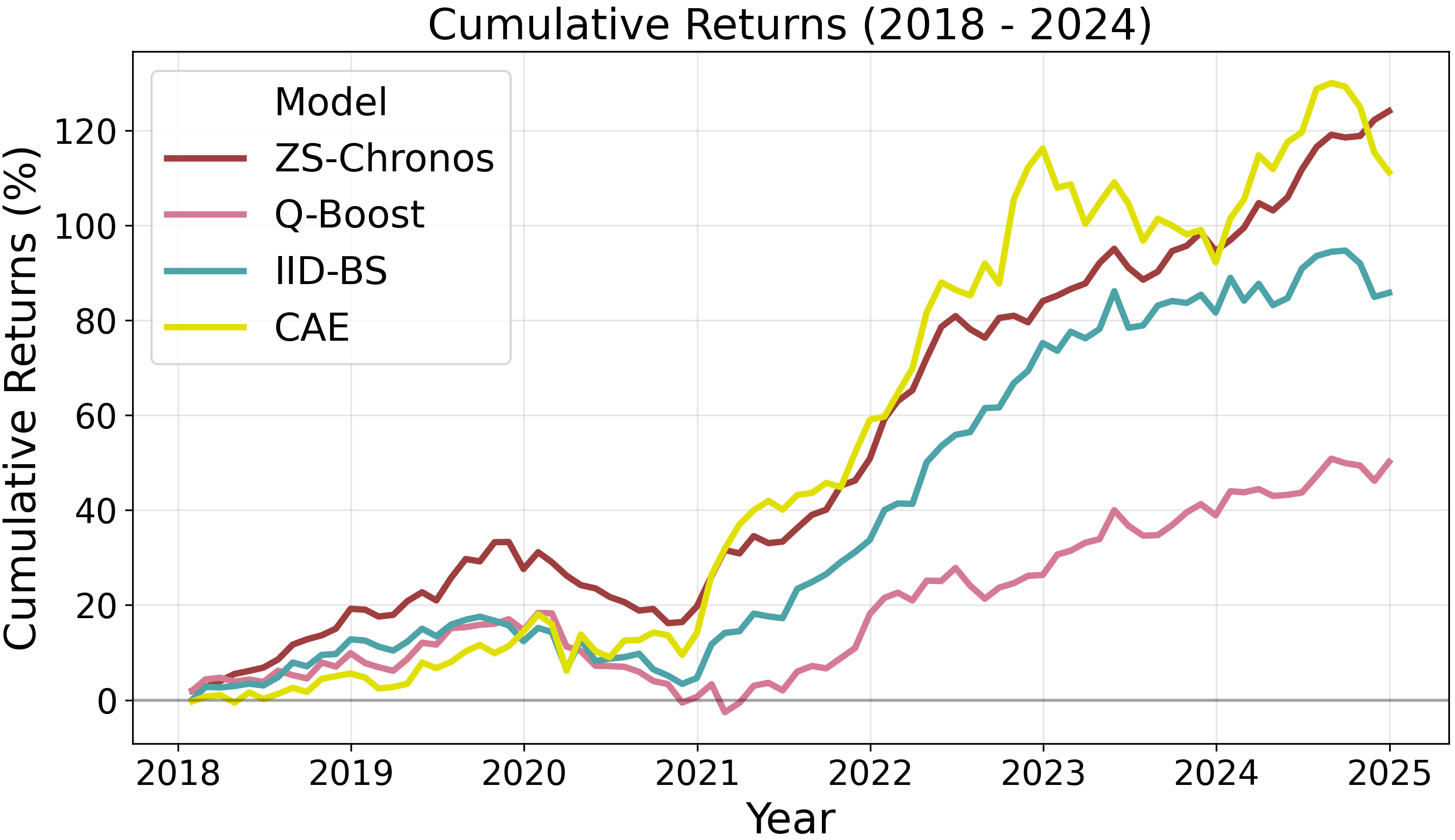}
    \caption*{(b) Model performance after dataset creation.}
    \caption{Comparison of model performance before and after dataset construction.}
    \label{fig:cumulative_returns_combined}
\end{figure}

Figure ~\ref{fig:cumulative_returns_combined} compares the performance of each forecasting model during two distinct periods: before and after the M4 dataset release, which was employed in the pretraining of the Chronos model. In the pre-2018 period, as illustrated by Table~\ref{tab:chronos_combined}, all models deliver comparable performance. This outcome indicates that ZS-Chronos does not possess any significant advantage over the other models before its exposure to the M4 dataset.

In contrast, a notable divergence in model performance becomes evident during the post-2018 period, as detailed in Table~\ref{tab:chronos_combined}. Specifically, ZS-Chronos attains the highest annual return (11.79\%) and demonstrates superior risk-adjusted metrics, achieving a 1.826 Sharpe ratio and a 3.898 Sortino ratio. Conversely, the other forecasting models exhibit relatively weaker performance, both in terms of absolute returns and risk-adjusted measures. If pretraining on the M4 dataset had introduced any forward-looking bias, ZS-Chronos would have been expected to display artificially inflated performance relative to other models in the pre-2018 period. Instead, the reverse is observed; ZS-Chronos significantly strengthens after the dataset’s creation.

This empirical evidence robustly supports the conclusion that ZS-Chronos’s superior post-2018 performance does not stem from data leakage or hidden pretraining biases. Importantly, the model is evaluated using synthetic latent factors specifically generated within our methodological framework, ensuring that Chronos had no prior exposure to the test inputs. Thus, the improved performance after 2018 further validates our approach, confirming that the pretrained Chronos model maintains strict data independence and achieves robust generalization. These findings substantiate the methodological rigor of employing pretrained foundation models such as Chronos within forecasting pipelines, emphasizing their reliability and practical applicability.

\section{Conclusions}

This paper introduces a scalable framework for high-dimensional CAEs in asset pricing, leveraging uncertainty-aware factor selection to mitigate performance degradation associated with increased latent dimensionality. Empirically, we show that selectively using latent factors based on forecast uncertainty consistently enhances risk-adjusted returns across multiple forecasting models. Notably, the highest performance emerges from portfolios that utilize only a subset of available latent factors.

We integrate three distinct forecasting methods—IID-BS, Q-Boost, and ZS-Chronos which are yielding largely uncorrelated predictive signals. This diversity enables ensemble strategies that significantly outperform individual models, with the best-performing ensemble achieving a $2.2$ Sharpe ratio, $4.01$ Sortino ratio, $5.95$ Omega ratio, and maximum drawdown below $10\%$. Furthermore, our adaptive factor selection, guided by temporally regularized log-sum-exp optimization, demonstrates robustness and practical relevance in realistic investment scenarios.

Beyond empirical validation, our framework provides a theoretical foundation for uncertainty-based selection. In particular, predictive uncertainty acts as a sufficient statistic for the expected degradation of portfolio utility under model misspecification. This interpretation links forecast dispersion directly to estimation risk and aligns with classical results showing that predictive variance governs expected utility loss and confidence weighting in optimal portfolios \cite{pastor2000portfolio, black1992global, avramov2010should}. By explicitly ranking factors according to their forecast uncertainty, the proposed approach operationalizes this principle to achieve robust out-of-sample efficiency.

Finally, our ablation analysis rules out potential data leakage concerns related to the pretrained ZS-Chronos model, confirming that its performance is not driven by hidden biases. By relying exclusively on internally generated synthetic latent factors, we preserve full data independence and maintain evaluation integrity. Overall, our findings establish uncertainty-driven dimensionality control as essential for effectively scaling CAE models, providing a principled solution to balancing the bias–variance tradeoff inherent in high-dimensional latent factor frameworks.

\bibliography{refs}

@article{neuhierl2022structural,
  author = {Neuhierl, Andreas and Varneskov, Rasmus T.},
  title = {Structural Breaks in Asset Pricing Models},
  journal = {Journal of Financial Economics},
  volume = {144},
  number = {1},
  pages = {290--315},
  year = {2022},
  publisher = {Elsevier}
}

@article{giglio2021asset,
  author = {Giglio, Stefano and Xiu, Dacheng},
  title = {Asset Pricing with Omitted Factors},
  journal = {Journal of Political Economy},
  volume = {129},
  number = {7},
  pages = {1947--1990},
  year = {2021},
  publisher = {University of Chicago Press}
}

@article{gu2020autoencoder,
  author = {Gu, Shihao and Kelly, Bryan and Xiu, Dacheng},
  title = {Autoencoder Asset Pricing Models},
  journal = {Journal of Econometrics},
  volume = {222},
  number = {1},
  pages = {429--450},
  year = {2020},
  publisher = {Elsevier}
}

@article{wei2025deeplatent,
  author = {Wei, Jason and Chen, Rui and Li, Bin},
  title = {Deep Latent Factor Model for Asset Pricing},
  journal = {Journal of Financial Data Science},
  volume = {7},
  number = {1},
  pages = {45--67},
  year = {2025}
}

@article{ansari2024chronos,
  author = {Ansari, Abdul Fatir and Stella, Lorenzo and Turkmen, Caner and Zhang, Xiyuan and Mercado, Pedro and Shen, Huibin and Shchur, Oleksandr and Rangapuram, Syama Sundar and Arango, Sebastian Pineda and Kapoor, Shubham and others},
  title = {Chronos: Learning the Language of Time Series},
  journal = {arXiv preprint arXiv:2403.07815},
  year = {2024}
}

@article{jensen2023replication,
  author = {Jensen, Theis Ingerslev and Kelly, Bryan and Pedersen, Lasse Heje},
  title = {Is There a Replication Crisis in Finance?},
  journal = {The Journal of Finance},
  volume = {78},
  number = {5},
  pages = {2465--2518},
  year = {2023},
  publisher = {Wiley Online Library}
}

@article{friedman2001greedy,
  author = {Friedman, Jerome H.},
  title = {Greedy Function Approximation: A Gradient Boosting Machine},
  journal = {Annals of Statistics},
  volume = {29},
  number = {5},
  pages = {1189--1232},
  year = {2001},
  publisher = {Institute of Mathematical Statistics}
}

@article{aksu2024gifteval,
  author = {Aksu, Taha and Chen, Gerald and Shi, Xinyu and Feng, Qihe and Zhang, Yan},
  title = {{GIFT-Eval}: A Benchmark for General Time Series Forecasting Model Evaluation},
  journal = {arXiv preprint arXiv:2410.10393},
  year = {2024}
}

@misc{rapids2018,
  author = {{NVIDIA Corporation}},
  title = {{RAPIDS}: {GPU} Accelerated Data Science},
  year = {2018},
  howpublished = {\url{https://rapids.ai}}
}

@article{pastor2000portfolio,
  author = {P\'{a}stor, \v{L}ubo\v{s}},
  title = {Portfolio Selection and Asset Pricing Models},
  journal = {The Journal of Finance},
  volume = {55},
  number = {1},
  pages = {179--223},
  year = {2000},
  publisher = {Wiley Online Library}
}

@article{black1992global,
  author = {Black, Fischer and Litterman, Robert},
  title = {Global Portfolio Optimization},
  journal = {Financial Analysts Journal},
  volume = {48},
  number = {5},
  pages = {28--43},
  year = {1992},
  publisher = {Taylor \& Francis}
}

@article{avramov2010should,
  author = {Avramov, Doron and Chordia, Tarun},
  title = {Should Investors Assimilate Return Predictability? {How} to Measure the Incremental Value of Conditioning Information},
  journal = {Journal of Financial Economics},
  volume = {98},
  number = {1},
  pages = {169--185},
  year = {2010},
  publisher = {Elsevier}
}

@article{Nesterov2005Smooth,
  author = {Nesterov, Yurii},
  title = {Smooth Minimization of Non-Smooth Functions},
  journal = {Mathematical Programming},
  volume = {103},
  number = {1},
  pages = {127--152},
  year = {2005},
  publisher = {Springer}
}

@inproceedings{McMahan2017FTRL,
  author = {McMahan, H. Brendan and Holt, Gary and Sculley, D. and Young, Michael and Ebner, Dietmar and Grady, Julian and Nie, Lan and Phillips, Todd and Davydov, Eugene and Golovin, Daniel and others},
  title = {Ad Click Prediction: A View from the Trenches},
  booktitle = {Proceedings of the 19th {ACM} {SIGKDD} International Conference on Knowledge Discovery and Data Mining},
  pages = {1222--1230},
  year = {2013},
  publisher = {ACM}
}

@article{Hazan2017Survey,
  author = {Hazan, Elad},
  title = {Introduction to Online Convex Optimization},
  journal = {Foundations and Trends in Optimization},
  volume = {2},
  number = {3-4},
  pages = {157--325},
  year = {2016},
  publisher = {Now Publishers, Inc.}
}

@article{Makridakis2020M4,
  author = {Makridakis, Spyros and Spiliotis, Evangelos and Assimakopoulos, Vassilios},
  title = {The {M4} Competition: 100,000 Time Series and 61 Forecasting Methods},
  journal = {International Journal of Forecasting},
  volume = {36},
  number = {1},
  pages = {54--74},
  year = {2020},
  publisher = {Elsevier}
}

@article{Makridakis2018M4,
  author = {Makridakis, Spyros and Spiliotis, Evangelos and Assimakopoulos, Vassilios},
  title = {The {M4} Competition: Results, Findings, Conclusion and Way Forward},
  journal = {International Journal of Forecasting},
  volume = {34},
  number = {4},
  pages = {802--808},
  year = {2018},
  publisher = {Elsevier}
}

\end{document}